\documentclass[a4paper,11pt]{article}
\usepackage{pos}
\usepackage{euscript}
\usepackage{fancybox}
\usepackage{enumerate}
\usepackage{widetext}
\usepackage{subfigure}
\usepackage[compat=1.0.0]{tikz-feynman}

\newcommand{\Meu}{\EuScript{M}}

\newcommand{\sfac}{\mathfrak{s}}

\def\st{\hbox{}} %% hbox is lower
\usepackage[figuresright]{rotating}
\title{Collinearly Enhanced YFS MC Approach to Precision High Energy Collider Physics}
\centerline{BU-HEPP-24-04, 09/2024}
\author*[a]{B.F.L. Ward}
\author[b]{S. Jadach\footnote[2]{Deceased.}} 
\author[b]{Z. Was}

\affiliation[a]{Department of Physics, Baylor University,\\
  Waco, TX, USA}

\affiliation[b]{Institute of Nuclear Physics,\\
Krakow, PL}

\emailAdd{bfl\_ward@baylor.edu}
\emailAdd{zbigniew.was@ifj.edu.pl}
\abstract{We improve the YFS IR resummation theory so that it includes all of the attendant collinear contributions which exponentiate. The attendant new resummed contributions are shown to agree with known results from the collinear factorization approach. We argue that they improve the corresponding precision tag for a given level of exactness in the respective YFS hard radiation residuals as the latter are realized in the YFS MC approach to precision high-energy collider physics.}

\FullConference{42nd International Conference on High Energy Physics (ICHEP2024)\\
18-24 July 2024\\
Prague, Czech Republic\\}

\begin{document}
\maketitle

\section{Introduction}
We present here the elements of the collinear improvement as given in Ref.~\cite{Jadach:2023cl1} of the original YFS algebra with an eye toward enhancing the physics expectations of future colliding beam devices as described in Refs.~\cite{Jadach:2019bye,frix-lnen:2022}. The starting point is the basic YFS formula for the cross-section under study, which we present for the prototypical process $e^+e^-\rightarrow f\bar{f}+n\gamma, \; f = \ell, q, \; \ell=e,\mu,\tau,\nu_e,\nu_\mu,\nu_\tau, \; q = u,d,s,c,b,t$ here as 
\begin{equation}
\sigma =\frac{1}{\text{flux}}\sum_{n=0}^{\infty}\int d\text{LIPS}_{n+2}\; \rho_A^{(n)}(\{p\},\{k\}),
\label{eqn-yfsmth-1}
\end{equation}
where $\text{LIPS}_{n+2}$ denotes Lorentz-invariant phase space for $n+2$ particles, $A=\text{CEEX},\;\text{EEX}$, the incoming and outgoing fermion momenta are abbreviated as 
$\{p\}$, and the $n$ photon momenta are denoted by $\{k\}$. We have (See Refs.~\cite{Jadach:2000ir,Jadach:1999vf}.)
\begin{equation}
\rho_{\text{CEEX}}^{(n)}(\{p\},\{k\})=\frac{1}{n!}e^{Y(\Omega;\{p\})}\bar{\Theta}(\Omega)\frac{1}{4}\sum_{\text{helicities}\;{\{\lambda\},\{\mu\}}}
\left|\Meu\left(\st^{\{p\}}_{\{\lambda\}}\st^{\{k\}}_{\{\mu\}}\right)\right|^2.
\label{eqn-yfsmth-2}
\end{equation}
(The corresponding formula for the $A=\text{EEX}$ case is given in Refs.~\cite{Jadach:2000ir,Jadach:1999vf}.) $Y(\Omega;\{p\})$ is the YFS infrared exponent.
The region $\Omega$ specifies the respective infrared integration limits and its characteristic function
$\Theta(\Omega,k)$ for a photon of energy $k$, with $\bar\Theta(\Omega;k)=1-\Theta(\Omega,k)$ and $$\bar\Theta(\Omega)=\prod_{i=1}^{n}\bar\Theta(\Omega,k_i).$$
By definition, $\Theta(\Omega,k)=1$  for $k\in\Omega$ 
and $\Theta(\Omega,k)=0$ for $k\not\in \Omega$.
We note that, for $\Omega$ defined with the condition $k^0<E_{\min},$ the YFS infrared exponent reads
%//////////////////////////////////////////////////
\begin{equation}
  \label{eq:YFS-ffactor}
  \begin{split}
   &Y(\Omega;p_1,...,p_4)
  =   Q_e^2   Y_\Omega(p_1,p_2)  +Q_f^2   Y_\Omega(p_3,p_4)\\
&\qquad\qquad
     +Q_e Q_f Y_\Omega(p_1,p_3)  +Q_e Q_f Y_\Omega(p_2,p_4) 
     -Q_e Q_f Y_\Omega(p_1,p_4)  -Q_e Q_f Y_\Omega(p_2,p_3),
  \end{split}
\end{equation}
where 
\begin{equation}
\label{eq:form-factor}
Y_\Omega(p,q) 
     \equiv  2 \alpha \tilde{B}(\Omega,p,q)   +2 \alpha \Re B(p,q) 
\end{equation}
is a sum of the real infrared contribution determined by the real emission infrared function $\tilde{B}$ and the virtual infrared contribution determined by the virtual infrared function $B$.
The latter infrared functions and the CEEX amplitudes ~$\{\Meu\}$ are defined in Refs.~\cite{Jadach:2000ir,Jadach:1999vf,Jadach:2013aha}. 
\section{Improving the Collinear Limit of YFS Theory}
The question naturally arises as to what, if any, non-soft\footnote{The soft collinear singularities are already included since the YFS soft limit is exact.} collinear singularities are resummed in the YFS resummation algebra carried by the YFS exponent $Y(\Omega;\{p\})$. We show in Ref.~\cite{Jadach:2023cl1} that in both the virtual and real contributions to $Y(\Omega;\{p\})$
it is possible to improve its collinear limit.\par
Specifically, using the virtual infrared contributions for definiteness, we find that the result in the s-channel for $Y(\Omega;\{p\})$ in eq.(\ref{eqn-yfsmth-2}), dropping terms  ${\cal O}(m^2/s)\; \text{where} \; s=(p_1+p_2)^2$, 
\begin{equation}
 \label{eq:form-factor1}
\begin{split}
  Y_e(\Omega_I;p_1,p_2)
 =   \gamma_e \ln {2E_{min}\over \sqrt{2p_1p_2}}     +{1\over 4}\gamma_e
      +Q_e^2 {\alpha\over\pi} \bigg( -{1\over 2} +{\pi^2\over 3}\bigg),
\end{split}
\end{equation}
with
\begin{equation}
\label{eq:form-factor2}
\gamma_e = 2 Q_e^2 {\alpha\over \pi} \bigg( \ln {2p_1p_2\over m_e^2} -1 \bigg),
\end{equation}
becomes, when the collinearly enhanced terms which exponentiate are retained, 
\begin{equation}
\begin{split}
  Y_{CL,e}(\Omega_I;p_1,p_2)
 =   \gamma_e \ln \frac{2E_{min}}{ \sqrt{2p_1p_2}}     +\frac{3}{4}\gamma_e
      +Q_e^2 \frac{\alpha}{\pi} \bigg( \frac{1}{ 2} \;+ \; \frac{\pi^2}{ 3}\bigg),
\end{split}
\end{equation}
using an obvious notation $CL$ to denote that the entire factor $\frac{3}{2}Q_e^2{\frac{\alpha}{\pi}} L$ is now exponentiated by our collinearly improved YFS virtual IR function $B_{CL}$ given
explicitly in Ref.~\cite{Jadach:2023cl1}. In eq.(\ref{eq:form-factor1}) only $\frac{1}{3}$ of the latter collinear big log term is exponentiated. \par
In Ref.~\cite{Jadach:2023cl1}, we show that a similar collinear enhancement can be realized in the real infrared function $\tilde{B}$ thereby improving it to its collinearly enhanced 
form $\tilde{B}_{CL}$ -- see Ref.~\cite{Jadach:2023cl1}.\par
The same methods allow us to obtain the collinearly enhanced form of the CEEX soft eikonal factors $\sfac_{\sigma}(k)\rightarrow \sfac_{CL,\sigma}(k)$ in an obvious notation
following that in Ref.~\cite{ceex2:1999}. See Ref.~\cite{Jadach:2023cl1} for further details.\par

%\begin{thebibliography}{99}
%\bibitem{cern-courier-2023}W. Placzek et al.
\bibliography{Tauola_interface_design}{}
\bibliographystyle{utphys_spires}
%\end{thebibliography}

\end{document}